\documentclass[preprint,12pt,3p]{elsarticle}

\usepackage{matlab-prettifier}
\usepackage{graphicx}
\usepackage{amsmath}
\usepackage{makecell}
\usepackage{xfrac}
\usepackage{parskip}
\usepackage[backref=page]{hyperref}
\usepackage{adjustbox}
\usepackage{setspace}
\hypersetup{ 
backref=true, 
bookmarks=true, 
pdftoolbar=true, 
pdfmenubar=true, 
pdffitwindow=false, 
pdfstartview={FitH}, 
pdftitle={}, 
pdfauthor={}, 
pdfsubject={}, 
pdfkeywords={}{}, 
pdfnewwindow=true, 
colorlinks=true, 
linkcolor=BlueViolet, 
citecolor=maroon, 
urlcolor=blue,
}
\usepackage{comment}
\usepackage{subfig}
\usepackage{caption}
\usepackage{longtable}
\usepackage{amssymb}
\usepackage{gensymb}
\usepackage{siunitx}

\usepackage{natbib}
\biboptions{comma,round}
\setcitestyle{authoryear}

\begin{document}
\begin{frontmatter}

\title{The social cost of carbon}

\author[label1,label2,label3,label4,label5,label6,label8]{Richard S.J. Tol\corref{cor1}}
\address[label1]{Department of Economics, University of Sussex, Falmer, United Kingdom}
\address[label2]{Institute for Environmental Studies, Vrije Universiteit, Amsterdam, The Netherlands}
\address[label3]{Department of Spatial Economics, Vrije Universiteit, Amsterdam, The Netherlands}
\address[label4]{Tinbergen Institute, Amsterdam, The Netherlands}
\address[label5]{CESifo, Munich, Germany}
\address[label6]{Payne Institute for Public Policy, Colorado School of Mines, Golden, CO, USA}
\address[label8]{College of Business, Abu Dhabi University, UAE}

\cortext[cor1]{Jubilee Building, BN1 9SL, UK}

\ead{r.tol@sussex.ac.uk}
\ead[url]{http://www.ae-info.org/ae/Member/Tol\_Richard}

\begin{abstract}
The social cost of carbon is the damage avoided by slightly reducing carbon dioxide emissions. It is a measure of the desired intensity of climate policy. The social cost of carbon is highly uncertain because of the long and complex cause-effect chain, and because it quantifies and aggregates impacts over a long period of time, affecting all people in a wide range of possible futures. Recent estimates are around \$80/tCO\textsubscript{2}.
\end{abstract}

\end{frontmatter}

The social cost of carbon is the damage done, at the margin, by emitting an additional tonne of carbon dioxide\textemdash or the benefit of reducing emissions by a little. Formally, the social cost of carbon in year $t$ is the first partial derivative of an intertemporal social welfare function to carbon dioxide emissions in year $t$, normalized by the marginal utility of consumption in year $t$. If evaluated along the corresponding optimal emissions trajectory, the social cost of carbon is the Pigou-Bator tax, the tax that internalizes the climate change externality.

The social cost of carbon is a key parameter in the economic analysis of climate policy \citep{Pizer2014, NAS2017, Pezzey2019}. It determines the price that, in a first-best world, should be put on greenhouse emissions\textemdash if the aim is to maximize social welfare. Together with the cost of greenhouse gas emission reduction, the social cost of carbon thus determines how much emissions should be cut and how much climate should change.

The social cost of carbon is important to policy too. A number of countries, including the USA, have an official social cost of carbon against which to evaluate public investment and regulation. Other countries, include the EU, have set targets for quantities rather than prices but are now beginning to realize that those targets require re-evaluation \citep{Tol2012EP}. 

While centrally important, the social cost of carbon is hard to estimate because there are so many moving parts in the analysis. This first partial derivative\textemdash the social cost of carbon\textemdash has a long chain rule:
\begin{itemize}
    \item Greenhouse gas emissions affect atmospheric concentrations.
    \item Excess concentrations cause radiative forcing of the atmosphere.
    \item Radiative forcing heats the atmosphere.
    \item Atmospheric warming sets in motion a series of feedback effects in the atmosphere, notably cloud formation and water vapour concentration, and on the planet's surface, notably ice and snow cover and vegetation, which in turn affect, mostly amplify, warming.
    \item Atmospheric warming is delayed by oceanic warming, which leads to sea level rise.
    \item Warming varies over space and between seasons, and affects patterns of rainfall, storminess, and a host of other weather variables.
    \item Climate change and sea level rise impact agriculture, energy use, energy supply, tourism and recreation, labour productivity, water resources, human health, air pollution, and biodiversity, among others.
    \item These impacts will be actively modified by people seeking to minimize the negative effects and make the most of the positive ones.
    \item As economies develop, technologies improve, and institutions are altered, the ability to cope with weather changes.  
    \item Residual impacts plus adaptation costs need to be expressed as a loss (or gain) of utility or the equivalent income loss.
\end{itemize}
If the long causal chain is not complicated enough, the monetized impacts occur at a particular time to a particular person in a particular scenario, and therefore need to be aggregated over space, time, and probability. After all, the social cost of carbon measures the externality, that is, the impact on all people.

None of the above steps is easy. All of the necessary parameters are uncertain, often very much so because the social cost of carbon is the marginal impact of \emph{future} climate change. 

Some assumptions are not just uncertain or ambiguous but controversial. Experts may disagree on mechanisms. Non-economists often criticize monetisation, while economists would fret about non-marginal changes, benefit transfer, and accuracy of stated preferences.

Aggregation of impacts is inherently problematic. A lot has been written about the ``correct'' discount rate. The literature remains divided between the acolytes of Aristotle, who argue for a particular discount rate from ethical or religious principles, and the disciples of Cleisthenes, who think that the discount rate should reflect the will of the people, while the followers of Baumol argue that the opportunity cost of capital is a better gauge of the price of time than the marginal value of future consumption. These camps argue amongst themselves and are challenged by those who say that \emph{exponential} discounting is mistaken\textemdash both as a description of how people make intertemporal trade-offs and as a prescription of how people should do this.

Other aspects of the intertemporal welfare function have attracted less attention but are important too for the social cost of carbon. Economists tend to agree on how to deal with risk but not on what the rate of risk aversion is or should be. Economists have less of a grip on how to handle fat tails, ambiguity, and stochasticity, all key features of climate change.

Comparing and adding the utilities of different people has to be done\textemdash this is unavoidable with a negative externality, as it imposes costs on a third party. Constructing representative agents or social welfare functions cannot be done in a satisfactory way.

Tied up in this is that the standard utility function is identically curved in time, risk, and equity space: Relative marginal utility is the same when we compare the same person at two points in time, the same person in two situations, or two persons. Separating impatience from risk aversion and from aversion to inequality between and within groups is a step forward that introduces yet more debatable parameter choices in the estimation of the social cost of carbon.

Furthermore, climate change affects mortality, fertility, and migration. There are no compelling intertemporal welfare functions with endogenous population growth. The impact on the social cost of carbon remains largely unexplored.

\begin{figure}
    \centering
    \includegraphics[width=\textwidth]{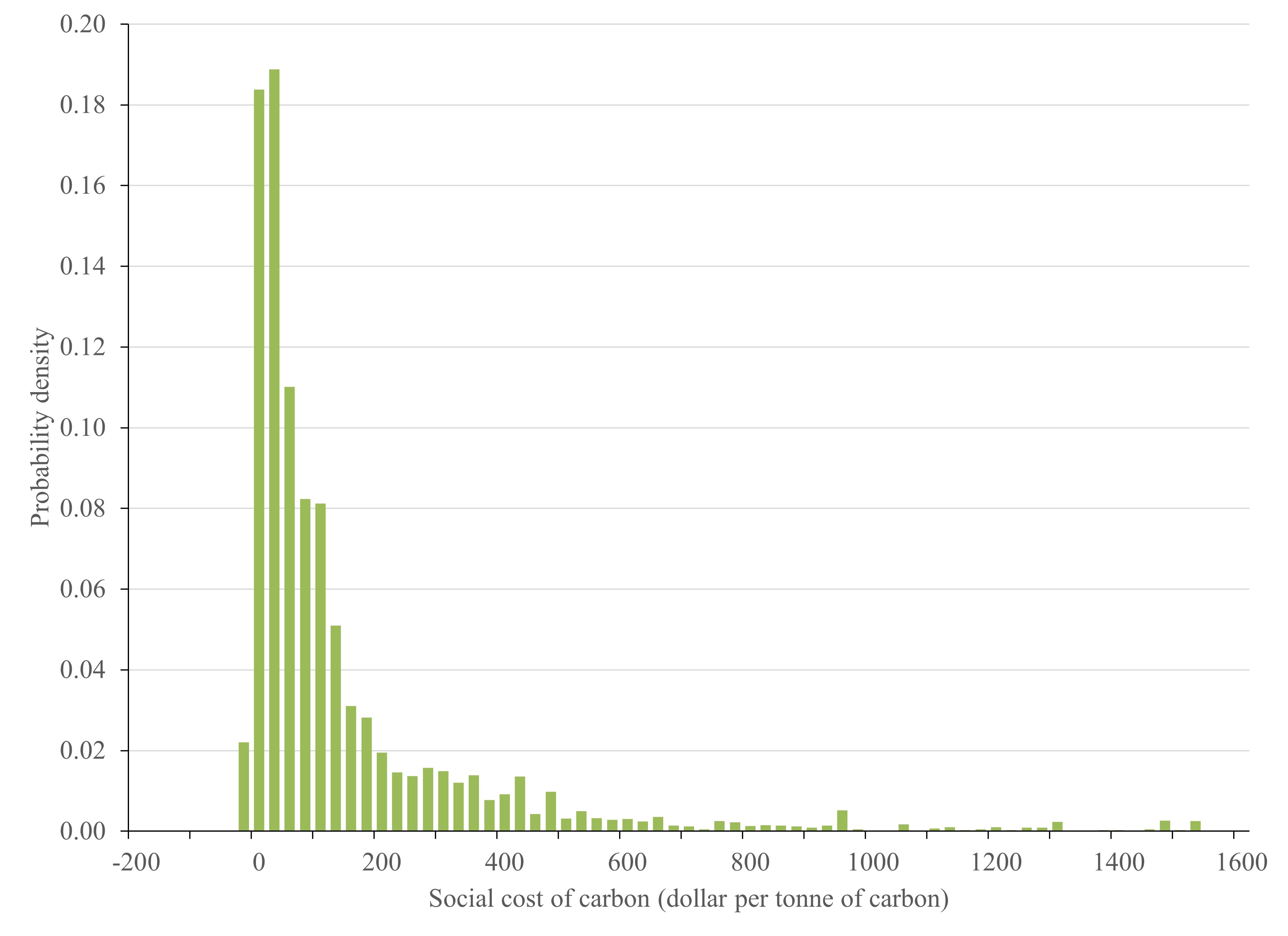}
    \caption{Histogram of the published estimates of the social cost of carbon, 1982-2022, quality-weighted and censored.}
    \caption*{\scriptsize The social cost of carbon is measured in 2010 U.S. dollar per metric tonne of carbon. Multiply by 12/44 in order to convert to a tonne of carbon dioxide. The social cost of carbon is for emissions in the year 2010.}
    \label{fig:histogram}
\end{figure}

Because it is both important and difficult to estimate, there is a large literature on the social cost of carbon. Figure \ref{fig:histogram} shows the histogram of 6,340 estimates published in 224 papers between \citet{Nordhaus1980} and 2022. Following \citet{Tol2023NCC}, the estimates are quality-weighted and censored. There is a small chance, 2.2\%, that the social cost of carbon is negative, that is, a benefit. The mode of the empirical distribution lies below \$50/tC but the mean is \$184/tC. Figure \ref{fig:histogram} reveals a pronounced skewness, with a thick right tail. The uncertainty is indeed large: The standard deviation is \$382/tC.

\begin{figure}
    \centering
    \includegraphics[width=\textwidth]{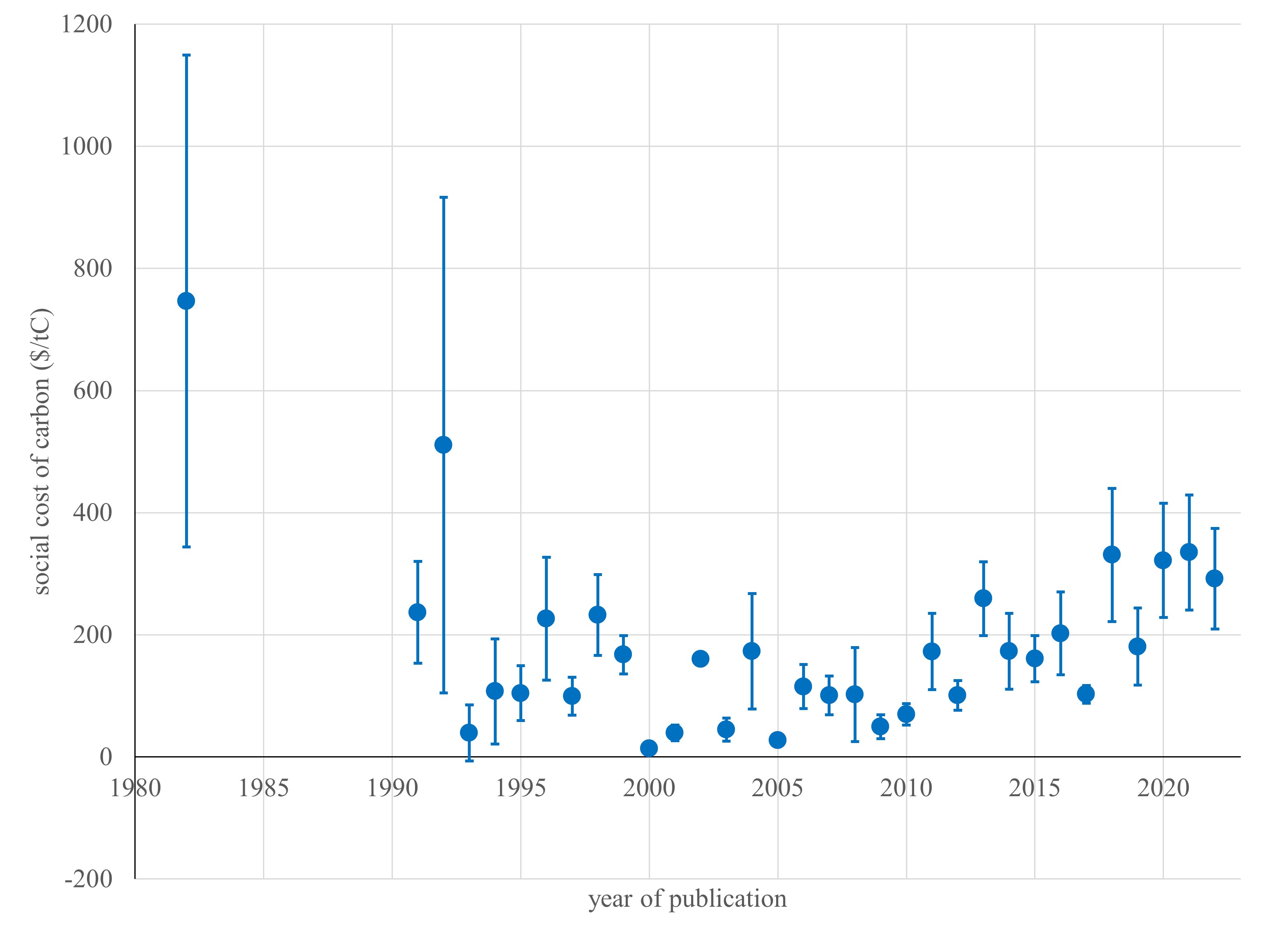}
    \caption{Annual mean of the published estimates of the social cost of carbon, quality-weighted and censored. Error bars are plus or minus the standard deviation.}
    \label{fig:time}
\end{figure}

Figure \ref{fig:time} shows the evolution over time. Early estimates of the social cost of carbon were very high, followed by a period of much lower numbers. Since the year 2000, there has been a steady increase in estimates of the social cost of carbon. This is partly because analysts tend to use a lower discount rate than they used to, but controlling for that, the upward trend is significant \citep{Tol2023NCC}.

\begin{figure}
    \centering
    \includegraphics[width=\textwidth]{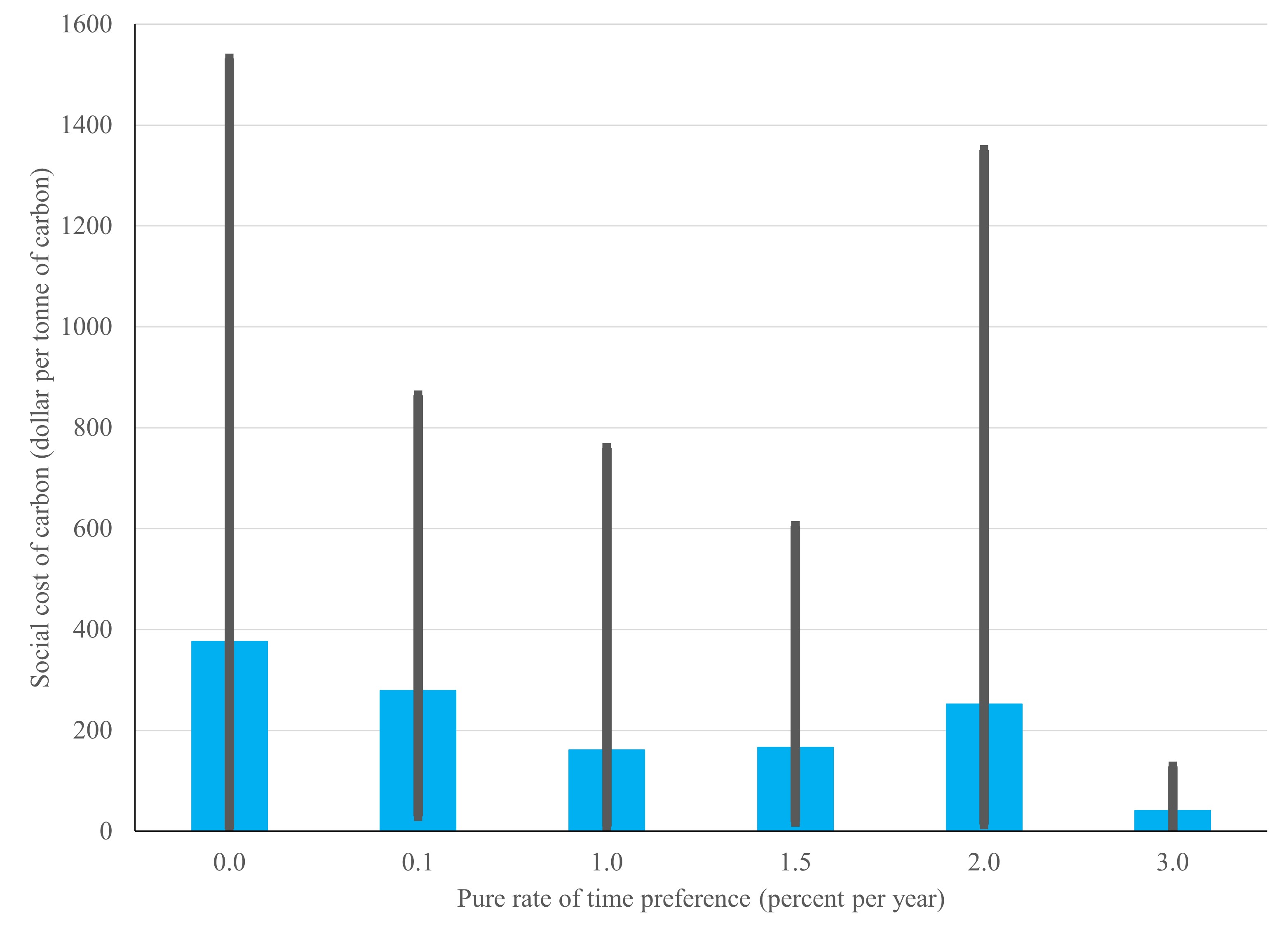}
    \caption{Mean of the published estimates of the social cost of carbon, quality-weighted and censored, by pure rate of time preference. Error bars are the 5th and 95th percentiles.}
    \label{fig:prtp}
\end{figure}

Estimates of the social cost of carbon differ for many reasons, but the choice of discount rate is a key difference. Figure \ref{fig:prtp} shows the mean and 90\% confidence interval for six popular choices of the pure rate of time preference. Ignoring the anomalous results for a utility discount rate of 2\% per year, the average estimate of the social cost of carbon increases as the discount rate falls. This is no surprise. The social cost of carbon measures the marginal loss of future welfare. The higher the discount rate, the less we care about the future and so about the impacts of climate change.

Figure \ref{fig:prtp} also shows that the uncertainty narrows as the discount rate increases. This is again as expected. The further we look into the future, the more uncertain things become. The pure rate of time preference therefore not only discounts the future, but also the uncertainties that come with that.

A \href{https://richardtol.shinyapps.io/MetaSCC/}{shiny app} \citep{TolTol2023} lets the reader look at other aspects of the same data.

A number of robust results can be found in the large literature. The social cost of carbon tends to increase if
\begin{itemize}
    \item emissions are higher;
    \item the climate sensitivity is higher;
    \item people are more numerous;
    \item the economy is larger;
    \item impacts are larger;
    \item vulnerability falls less fast;
    \item the discount rate is lower;
    \item risk aversion is larger; or
    \item inequity aversion is stronger.
\end{itemize}
This is because impacts are more than linear in climate change; because impacts are in the future, are uncertain, and disproportionally affect poorer people; and because the social cost of carbon is an absolute number. I write ``tends to'' because there are exceptions to all these points, but you would need to torture your integrated assessment model to find those. 

The numerous papers on the social cost of carbon notwithstanding, much research remains to be done. For instance, the majority of the estimates of the total impact of climate change \citep{Tol2023meta} have never been used to assess the social cost of carbon. Faster computers allow for more complex integrated assessment models, and models with a finer spatial and temporal resolution. More and better data create new opportunities for the validation of calibrated models, and the construction of estimated models. Increased research capacity means that the views and perspectives of people outside the global north can be better represented. New and, more importantly, different estimates of the social cost of carbon will emerge, along with new insights into what is and what is not important. For the moment, the data in Figure \ref{fig:histogram} is the best we have.

\bibliography{master}

\end{document}